\documentstyle[prl,aps,epsf,amsmath,amssymb]{revtex}
\begin{document}
\draft \title{An Elementary Proof of Lyapunov Exponent Pairing for
Hard-Sphere Systems at Constant Kinetic Energy} \author{Debabrata
Panja\\} \maketitle
\begin{center}
{\em Instituut-Lorentz, Universiteit Leiden, Postbus 9506, 2300 RA
Leiden, The Netherlands. \/}
\end{center}
\begin{abstract}

\noindent The conjugate pairing of Lyapunov exponents for a
field-driven system with smooth inter-particle interaction at constant
total kinetic energy was first proved by Dettmann and Morriss
[Phys. Rev. E {\bf 53}, R5545 (1996)] using simple methods of
geometry. Their proof was extended to systems interacting via
hard-core inter-particle potentials by Wojtkowski and Liverani
[Comm. Math. Phys. {\bf 194}, 47 (1998)], using more sophisticated
methods. Another, and somewhat more direct version of the proof for
hard-sphere systems has been provided by Ruelle [J. Stat. Phys. {\bf
95}, 393 (1999)]. However, these approaches for hard-sphere systems
are somewhat difficult to follow. In this paper, a proof of the
pairing of Lyapunov exponents for hard-sphere systems at constant
kinetic energy is presented, based on a very simple explicit geometric
construction, similar to that of Ruelle. Generalizations of this
construction to higher dimensions and arbitrary shapes of scatterers
or particles are trivial. This construction also works for hard-sphere
systems in an external field with a Nos\'e-Hoover thermostat. However,
there are situations of physical interest, where these proofs of
conjugate pairing rule for systems interacting via hard-core
inter-particle potentials break down.
\end{abstract}
\vspace{4mm}

Keywords: Lorentz gas, Hard-sphere systems, Lyapunov exponents,
Conjugate pairing rule

\section{Introduction}

Thermostatted, field-driven systems have been popular models for
non-equilibrium molecular dynamics (NEMD) simulation studies of
transport processes in fluids. NEMD studies consider systems with a
large number of particles interacting with each other, driven by an
external field \cite{EM_book_academic,Hoover_book_elsevier}. In these
studies, the thermostat  continuously removes the energy generated
inside the system due to the work done on it by the external field, by
means of a dynamical friction term in the equations of motion. One
finds that for these systems, a non-equilibrium steady state (NESS),
homogeneous in space, is reached after a sufficiently long time
\cite{HLM_prl_82,Evans_jcp_83,Nose_mp_84,Nose_jcp_84,Hoover_pra_85,Morriss_pla_88}.
Here we consider a particular kind of thermostat, where the friction
is linearly coupled to the laboratory momenta and keeps the total
laboratory kinetic energy of the particles constant (i.e, isokinetic
Gaussian thermostat coupled to the laboratory momenta). The dynamical
description of these systems, would be Hamiltonian in the absence of
this thermostat, and the corresponding Hamiltonian system, obtained by
dropping the dynamical friction term from the equations of motion,
will be referred to as the {\em background Hamiltonian system}. Due to
the presence of the dynamical friction terms in the equations of
motion, such a system is no longer Hamiltonian with a conservation of
phase space volumes, but instead is phase-space contracting
\cite{EM_book_academic,D_book_cup}. The sum of all the Lyapunov
exponents, which measures the rate of long-time exponential growth of
phase-space volume, for such a system is thus negative
\cite{D_book_cup}. Of course, the background Hamiltonian system, if
chaotic, is such that the Lyapunov exponents sum up to
zero. Furthermore, due to its symplectic form, the Lyapunov exponents
of the background system come in pairs such that the sum of each such
pair is also zero \cite{Arnold_book_Springer,AM_book_Benjamin}. The
phenomenon of such pairing of Lyapunov exponents, where the sum of
each pair of non-zero exponents takes a constant value independent of
the particular pair, is known as the {\em conjugate pairing rule}
(CPR).  Dettmann and Morriss \cite{DM_pre_96} have studied the
isokinetic Gaussian thermostatted field-driven systems that are under
consideration here, assuming that the particles of the system interact
with smooth pair potential energies, and that the forces on the
particles due to the external field depends {\it only on their
positions}. They have proved that under these conditions, in a
restricted subspace of the phase space of all the particles,
characterized by all the non-zero Lyapunov exponents, such a system is
$\mu$-symplectic\footnote{For a definition of $\mu$-symplecticity
condition, see Eqs. (\ref{e7a}) and (\ref{e7b}) of this paper. The
usual symplectic condition is a particular case  of Eq. (\ref{e7a})
with $\mu\,=\,1$.}. As a consequence, the CPR is exactly satisfied in
that subspace, and it is independent of the number of particles in the
system (corresponding simulation results can be found in
\cite{DMR_pre_95,SEI_chaos_98}) --- the sum of each pair comes out to
be the same negative constant. One important consequence of this
result is that the macroscopic transport coefficients of these
systems, in the linear order, can be obtained from this constant value
of the sum (see, for example, \cite{SEI_mcs_98}). The restricted
subspace, characterized by all the non-zero Lyapunov exponents, is
identified by observing that with an isokinetic Gaussian thermostat,
trajectories of the system always lie on a constant total kinetic
energy hypersurface in the phase space of all the particles. This
constraint generates a zero Lyapunov exponent. Also, two points in the
phase space do not separate exponentially in time if one point is
chosen in the direction of flow of the other. This generates another
zero Lyapunov exponent.

As hard-core inter-particle interaction can be dealt as a limiting
case of a very short range smooth potential, one would expect that in
this restricted subspace, the system is still $\mu$-symplectic, and as
a result, the CPR will continue to hold for a field-driven isokinetic
Gaussian thermostatted system with hard-core inter-particle
interactions. The corresponding pairing of Lyapunov exponents for
hard-sphere systems has been proved by direct means using the
differential geometric structure of the phase space, by Wojtkowski and
Liverani \cite{WL_cmp_98}. Another version of the proof for
hard-sphere systems has been obtained by Ruelle
\cite{Ruelle_jsp_99}. The above two approaches for hard-sphere systems
are somewhat difficult to follow. The purpose of this paper therefore,
is to present a proof of the pairing of Lyapunov exponents (also by
direct means) for hard-sphere systems at constant kinetic energy,
based on a very simple explicit geometric construction, similar to
that used by Ruelle. Two kinds of field-driven isokinetic Gaussian
thermostatted systems with hard-core interaction are considered here:
in section II, the proof of the CPR is carried out for the
three-dimensional Lorentz gas, where mutually non-interacting point
particles suffer specular collisions with fixed spherical
scatterers. In section III, the proof is then carried out for a gas of
hard spheres. The explicit method of this paper allows one to identify
the dependence of these approaches (described in
Refs. \cite{WL_cmp_98,Ruelle_jsp_99} and here) on the geometrical
shapes of the scatterers (or the particles, as the case may be) and on
the nature of the externally applied field. Based on it,
generalizations to higher dimensions, to arbitrary geometry of the
scatterers or the particles, and to the case where the masses of the
particles are arbitrary, become trivial. Finally, in Sec. IV, we argue
that this construction can be used to prove the CPR for hard-sphere
systems in an external field with a Nos\'e-Hoover thermostat. We also
identify situations of physical interest, where these approaches break
down.

\section{Proof of the CPR for three-dimensional Lorentz gas}

The Lorentz gas model consists of a set of scatterers fixed in space
together with mutually non-interacting moving particles that suffer
elastic, specular collisions with the scatterers. Here we consider the
version of the model in three dimensions where the scatterers are hard
spheres, and are placed in space without overlapping. Each of the
moving particles is a point particle with unit mass $(m\,=\,1)$, and
is subjected to an external force that depends {\it only on its
position}, as well as a Gaussian thermostat which is designed to keep
its kinetic energy at a constant value.  During a flight, the equation
of motion of a particle is
\begin{eqnarray}
\dot{\bf
\Gamma}\,=\,\left[\,\dot{\vec{r}},\,\,\,\dot{\vec{p}}\,\right]\,=\,\left[\,\vec{p}\,,\,\,\,\vec{F}\,-\,\alpha\vec{p}\,\right]\,,
\label{e1}
\end{eqnarray}
where $\vec{F}\,=\,-\,\vec{\nabla}\phi$ is the force on the particle
due to the external field and $\alpha$ is the coefficient of dynamical
friction representing the isokinetic Gaussian thermostat. The value of
$\alpha$ is obtained from the fact that the kinetic energy of the
particle $\displaystyle{\frac{p^{2}}{2}}$ is constant during a flight,
i.e,
\begin{eqnarray}
\alpha\,=\,\frac{\vec{F}\cdot{\vec{p}}}{p^{2}}\,
\label{e2}
\end{eqnarray}
and without any loss of generality, the time hereafter is rescaled
such that $p^{2}\,=\,1$. At a collision with a scatterer, the
post-collisional position and momentum of a particle, ${\bf
\Gamma}_{+}\,=\,[\vec{r}_{+},\,\vec{p}_{+}]$, are related to its
pre-collisional position and momentum
${\bf\Gamma}_{-}\,=\,[\vec{r}_{-},\,\vec{p}_{-}]$, by
\begin{eqnarray}
{\bf \Gamma}_{+}\,=\,{\bf Q}\,{\bf
\Gamma}_{-}\,=\,\left[\,\vec{r}_{-}\,,\,\,\,\vec{p}_{-}\,-\,2\,(\vec{p}_{-}\cdot\hat{n})\,\hat{n}\,\right]\,,
\label{e3}
\end{eqnarray}
where $\hat{n}$ is the unit vector from the center of the scatterer to
the point of collision (see Fig. 1).

It is sufficient to consider the motion of only one of the moving
particles in its six-dimensional phase space to investigate the
chaotic properties of this system, since they do not interact with
each other. To obtain the Lyapunov exponents, we consider a particle
at the phase space location
${\bf\Gamma}_0\equiv[\vec{r}_{0},\,\vec{p}_{0}]$ at time $t=0$. In
time $t$, it suffers $s$ sequential collisions at time instants
$t_{1},t_{2},\ldots,t_{s}$ with the scatterers. In between collisions,
the particle undergoes flights, acted upon by the external field. We
refer to the phase space trajectory of  it as the ``reference
trajectory''. We also consider another particle at
${\bf\Gamma}_0+\delta{\bf\Gamma}_0\equiv[\vec{r}_{0}\,+\,\vec{\delta
r}_{0},\,\vec{p}_{0}\,+\,\vec{\delta p}_{0}]$ at time $t=0$, such that
${\bf\Gamma}_0$ and ${\bf\Gamma}_0+\delta{\bf\Gamma}_0$ are
infinitesimally apart from each other. This particle suffers the same
sequence of collisions and likewise its movement in the phase space
generates the ``adjacent trajectory''. The two trajectories  remain
infinitesimally apart from each other at all times. Let ${\bf
H}(t_{j}\,-\,t_{j-1})$ denote the time evolution operator for
$\delta{\bf\Gamma}(t)$ due to a flight between time instants $t_{j-1}$
and $t_{j}$ and let ${\bf M}_{i}$ denote the evolution operator for
$\delta{\bf\Gamma}(t)$ at the $i$-th collision. For
$0<t_1<t_2<\ldots<t_s<t$, we therefore have
\begin{eqnarray}
\delta{\bf\Gamma}(t)\,=\,{\bf H}(t\,-\,t_{s})\,{\bf M}_{s}\,{\bf
H}(t_{s}\,-\,t_{s-1})\,.\,.\,.\,.\,.\,.\,{\bf M}_{1}\,{\bf
H}(t_{1})\,\delta{\bf\Gamma}_0\,=\,{\bf L}(t)\,\delta{\bf\Gamma}_0\,.
\label{e4}
\end{eqnarray}
Let us now also define another six-dimensional matrix ${\bf T}(t)$,
such that during a flight
\begin{eqnarray}
\dot{\delta{\bf\Gamma}}(t)\,=\,{\bf T}(t)\,\delta{\bf\Gamma}(t)\,.
\label{e5}
\end{eqnarray}
The matrix ${\bf L}(t)$ can then be obtained from the solution of the
differential equation
\begin{eqnarray}
\dot{\bf H}(t)\,=\,{\bf T}(t)\,{\bf H}(t)\,,
\label{e6}
\end{eqnarray}
and Eq. (\ref{e4}), with the boundary condition that ${\bf
H}(0)\,=\,{\bf L}(0)\,=\,{\bf I}\,$. The Lyapunov exponents, measuring
the rate of exponential separation between the reference point and
adjacent point in this six-dimensional phase space for long times, are
then defined as the logarithms of the eigenvalues of the matrix ${\bf
\Lambda}$, where
\begin{eqnarray}
{\bf \Lambda}\,=\,\lim_{t\rightarrow\infty}\{[{\bf
L}(t)]^{\mbox{\scriptsize T}}{\bf L}(t)\}^{1/2t}\,.
\label{e7}
\end{eqnarray}

Clearly, there can be at most six Lyapunov exponents of this system,
two of which are zero due to the reasons explained in the
introduction. For our purpose, we need to select out a four-dimensional
subspace of this six-dimensional phase space where all the Lyapunov
exponents may be non-zero. In this four-dimensional subspace, the
proof of the CPR would follow from the $\mu$-symplecticity property of
${\bf L}(t)$ \cite{DM_pre_96,WL_cmp_98,Ruelle_jsp_99}. Thus, all we
need to prove to establish the CPR in this four-dimensional subspace
is that there exists a $\mu(t)$, such that with the four-dimensional
subspace representation of ${\bf L}(t)$,
\begin{eqnarray}
\mu(t)\,[{\bf L}(t)]^{\,\scriptsize{\mbox T}}\,{\bf J}\,[{\bf
L}(t)]\,=\,{\bf J}\,,
\label{e7a}
\end{eqnarray}
which is the $\mu$-symplecticity condition. Here, ${\bf J}$ is a
$4\times 4$ matrix with each entry being a $2\times 2$ matrix :
\begin{eqnarray}
{\bf J}\,=\,\left[\begin{array}{cc}\,\,\,{\bf
0}\,\,\,\,\,\,\,\,\,\,\,\,{\bf I}\\-\,{\bf I}\,\,\,\,\,\,\,\,\,\,{\bf
0}\end{array}\right]\,.
\label{e7b}
\end{eqnarray}
Our goal is to prove Eq. ({\ref{e7a}}) by means of the geometric
construction, developed in Ref. \cite{DM_pre_96}.

The identification of this four-dimensional subspace is facilitated by
decomposing the six-dimensional position and momentum space of the
particle at every point on the reference trajectory, into two separate
three-dimensional subspaces, one for the position-space and the other
for the momentum-space. Next, in each of these two three-dimensional
subspaces, a unit vector $\hat{e}_{0}\,=\,\vec{p}\,$ is chosen and two
other unit vectors $\hat{e}_{1}$ and $\hat{e}_{2}$ are also chosen to
form a complete set of orthonormal basis.  One zero Lyapunov exponent,
which occurs due to the fact that $\vec{\delta p}$ must be chosen
orthogonal to $\hat{e}_{0}$ in order to respect the constraint that
$p^{2}\,=\,1$, is avoided by measuring $\vec{\delta p}$ by its
components along the local directions of $\hat{e}_{1}$ and
$\hat{e}_{2}$, i.e,
\begin{eqnarray}
\vec{\delta p}\,=\,\sum_{i\,=\,1}^{2}\,\delta p_{i}\,\hat{e}_{i}\,.
\label{e7c}
\end{eqnarray}
The other zero Lyapunov exponent, which occurs due to the fact that
the adjacent point does not exponentially separate from the reference
point if $\vec{\delta r}_0$ is chosen along $\hat{e}_{0}(t={0})$, is
avoided by measuring $\vec{\delta r}$ also by its components along the
local directions of $\hat{e}_{1}$ and $\hat{e}_{2}$, i.e,
\begin{eqnarray}
\vec{\delta r}\,=\,\sum_{i\,=\,1}^{2}\,\delta r_{i}\,\hat{e}_{i}\,.
\label{e7d}
\end{eqnarray}
Albeit $\hat{e}_{0}$, being the momentum of the particle, is uniquely
defined at each point on the reference trajectory, $\hat{e}_{1}$ and
$\hat{e}_{2}$ can be chosen arbitrarily at every point of the
reference trajectory, maintaining the orthonormality condition. This
ambiguity in the local orientations of $\hat{e}_{1}$ and $\hat{e}_{2}$
can be removed by an initial choice of $\hat{e}_{1}(t={0})$ and
$\hat{e}_{2}(t={0})$ at $(\vec{r}_0, \vec{p}_{0})$ and subsequently
connecting the orthonormal set of basis vectors at different points on
the reference trajectory by means of a ``parallel transport''. The
construction and parallel transport of these basis vectors are the key
components of the proof of the $\mu$-symplecticity in this procedure.

To this end, we first concentrate on the evolution of
$\delta{\bf\Gamma}(t)$ during the process of a flight of the particle,
which is a special case of the systems that Dettmann and Morriss
\cite{DM_pre_96} have considered, where $\phi$ is the potential due to
{\it only\/} the external field. Following their construction, we
therefore use
\begin{eqnarray}
\dot{\hat{e}}_{i}\,=\,-\,(\vec{F}\cdot\hat{e}_{i})\,\hat{e}_{0},\hspace{1cm}i\,=\,1,2
\label{e8}
\end{eqnarray}
while the parallel transport of $\hat{e}_{0}$ is obtained from the
equations of motion, Eq. (\ref{e1}), i.e,
\begin{eqnarray}
\dot{\hat{e}}_{0}\,=\,\sum_{i\,=\,1}^{2}\,(\vec{F}\cdot\hat{e}_{i})\,\hat{e}_{i}\,.
\label{e9}
\end{eqnarray}
From Eq. (\ref{e1}), having obtained
\begin{eqnarray}
\dot{\bf
\Gamma}\,=\,\left[\,\hat{e}_{0}\,,\,\,\,{\displaystyle{\sum_{i\,=\,1}^{2}}}(\vec{F}\cdot\hat{e}_{i})\,\hat{e}_{i}\,\right]\,,
\label{e9a}
\end{eqnarray}
it is then easy to show that during a flight
\begin{eqnarray}
\dot{\delta r}_{i}\,=\,\delta
p_{i}\quad\quad\mbox{and}\quad\quad\dot{\delta
p}_{i}=\!\!\sum_{j\,=\,1}^{2}[-\,\nabla_{i}\nabla_{j}\phi\,-(\vec{F}\cdot\hat{e}_{i})(\vec{F}\cdot\hat{e}_{j})]\,\delta
r_{j}-\alpha\,\delta p_{i}\,.
\label{e10}
\end{eqnarray}
Consequently, the matrix ${\bf T}$ can be expressed as
\begin{eqnarray}
{\bf T}\,=\,\left[\begin{array}{cc}{\bf
0}\,\,\,\,\,\,\,\,\,\,\,\,\,\,\,\,\,\,{\bf I}\\{\bf
R}_{f}\,\,\,\,-\alpha{\bf I}\end{array}\right]\,,
\label{e12}
\end{eqnarray}
where each of the elements in ${\bf T}$ above is a $2\times 2$ matrix
and ${\bf R}_{f}$ is a symmetric matrix. The matrix ${\bf T}$ in
Eq. (\ref{e12}) has the property that
\begin{eqnarray}
{\bf T}^{ \scriptsize{\mbox T}}{\bf J}\,+\,{\bf J}{\bf
T}\,=\,-\,\alpha\,{\bf J}\,,
\label{e13}
\end{eqnarray}
from which it can be easily shown that for a flight between
$t_{j\,-\,1}$ and $t_{j}$
\begin{eqnarray}
\mu(t_{j}\,-\,t_{j-1})\,[{\bf
H}(t_{j}\,-\,t_{j-1})]^{\,\scriptsize{\mbox T}}\,{\bf J}\,[{\bf
H}(t_{j}\,-\,t_{j-1})]\,=\,{\bf J}\,,
\label{e15}
\end{eqnarray}
where
$\mu(t_{j}\,-\,t_{j-1})\,=\,\exp\left[\int_{t_{j-1}}^{t_{j}}\alpha(t')dt'\right]$
along the reference trajectory.

Next, we study the evolution of the infinitesimal volume element
$\delta{\bf\Gamma}$ due to a collision with a scatterer to obtain a
similar four-dimensional representation of the matrix ${\bf M}$,
defined in Eq. (\ref{e4}). We notice that even though Eqs. (\ref{e8})
and (\ref{e9}) uniquely determine the orientations of the basis
vectors over a flight given their orientation at the initiation of the
flight, the equations of parallel transport connecting the
pre-collisional and post-collisional basis vectors are still
lacking. As the  post-collisional momentum of the particle $\hat{e}_{0
+}$ can be related to its pre-collisional momentum $\hat{e}_{0 -}$ by
\begin{eqnarray}
\hat{e}_{0 +}\,=\,\hat{e}_{0 -}\,-\,2\,(\hat{e}_{0
-}\cdot\hat{n})\,\hat{n}\,,
\label{e16}
\end{eqnarray}
over a collision [see Eq.  (\ref{e3})], a parallel transport of the
orthonormal basis vectors that serves our purpose, can be completed
also by using
\begin{eqnarray}
\hat{e}_{i +}\,=\,\hat{e}_{i -}\,-\,2\,(\hat{e}_{i
-}\cdot\hat{n})\,\hat{n}\,.\hspace{1cm}i\,=\,1,\,2
\label{e17}
\end{eqnarray}
With the use of Eqs. (\ref{e7c}-\ref{e7d}) and (\ref{e16}-\ref{e17}),
we now show that as the infinitesimal pre-collisional phase space
separation can be written as
\begin{eqnarray}
{\bf \delta\Gamma}_{-}\,=\,\left[\,\vec{\delta
r}_{-}\,,\,\,\,\vec{\delta p}_{-}\,\right]
\,=\,\left[\,\displaystyle{\sum_{i\,=\,1}^{2}}\,\delta
r_{i-}\,\hat{e}_{i-}\,,\,\,\,\displaystyle{\sum_{i\,=\,1}^{2}}\,\delta
p_{i-}\,\hat{e}_{i-}\,\right]\,,
\label{e18}
\end{eqnarray}
the corresponding infinitesimal post-collisional separation,
$\delta{\bf \Gamma}_{+}$, can then be expressed as
\begin{eqnarray}
{\bf \delta\Gamma}_{+}\,=\,\left[\,\vec{\delta
r}_{+}\,,\,\,\,\vec{\delta
p}_{+}\,\right]\,=\,\left[\,\displaystyle{\sum_{i\,=\,1}^{2}}\,\delta
r_{i+}\,\hat{e}_{i+}\,,\,\,\,\displaystyle{\sum_{i\,=\,1}^{2}}\,\delta
p_{i+}\,\hat{e}_{i+}\,\right]\,,
\label{e19}
\end{eqnarray}
which would also reduce the collision dynamics of $\delta{\bf
\Gamma}(t)$ to four dimensions. We want to obtain the symplectic
property of the $4\times 4$ matrix ${\bf M}$, defined by
\begin{eqnarray}
{\bf \delta\Gamma}_{+}\,=\,{\bf M}\,{\bf \delta\Gamma}_{-}\,.
\label{e20}
\end{eqnarray}

We begin by studying Fig. 2, which is the two-dimensional projection
of an exaggerated picture of a collision that is taking place in
three-dimensions. The plane $\Delta_{-}$, perpendicular to the
reference trajectory at its point of collision A intersects the
adjacent trajectory at B. While $\vec{\delta r}_{-}\,=\,\vec{AB}$,
$\vec{\delta p}_{-}$ too lies on the plane $\Delta_{-}$. Similarly,
$\Delta_{+}$ is the plane that is perpendicular to the reference
trajectory at D passing through C, the point of collision of the
adjacent trajectory. The post-collisional position-space separation
between the two trajectories, $\vec{\delta r}_{+}\,=\,\vec{DC}$ and
$\vec{\delta p}_{+}$ also lies on the plane $\Delta_{+}$.  Clearly,
there is a time gap $\delta\tau$ between the two collisions at A and C
(i.e, the time required for the adjacent point to travel from B to C),
given by
\begin{eqnarray}
\delta\tau\,=\,-\,\frac{\vec{\delta r}_{-}\cdot\hat{n}}{\hat{e}_{0
-}\cdot\hat{n}}\,.
\label{e21}
\end{eqnarray}
Following the procedure outlined in \cite{DPH_pre_96}, we find that
the infinitesimal phase space separation of the two trajectories just
before the two collisions at A and C is
\begin{eqnarray}
\delta{\bf\Gamma}^{*}\,=\,\,{\bf
\delta\Gamma}_{-}\,+\,\dot{\bf\Gamma}_{-}\,\delta\tau
\label{e21a}
\end{eqnarray}
and consequently,
\begin{eqnarray}
{\bf \delta\Gamma}_{+}\,=\,\frac{\partial{\bf Q}}{\partial{\bf
\Gamma}_{-}}\cdot\delta{\bf\Gamma}^{*}\,-\,\dot{\bf\Gamma}_{+}\,\delta\tau\,
\label{e22}
\end{eqnarray}
where $\dot{\bf\Gamma}_{-}\,(\dot{\bf\Gamma}_{+})$ is obtained from
the equations of motion of the particle for the flight immediately
before and after the collision at A [see Eq. (\ref{e9a})], i.e,
\begin{eqnarray}
\dot{\bf
\Gamma}_{\pm}\,=\,\left[\,\hat{e}_{0\pm}\,,\,\,\,\displaystyle{\sum_{i\,=\,1}^{2}}(\vec{F}^{*}\cdot\hat{e}_{i\pm})\,\hat{e}_{i\pm}\,\right]
\label{e23}
\end{eqnarray}
($\vec{F}^{*}$ is the force on the reference point due to the external
field at the point of collision at A) and
\begin{eqnarray}
\frac{\partial{\bf Q}}{\partial{\bf
\Gamma}_{-}}\,=\!\left[\begin{array}{cc}{\,\,\,\,\,\,\,\,\,\,\,\,\,\,\,\,\bf
I}\,\,\,\,\,\,\,\,\,\,\,\,\,\,\,\,\,\,\,\,\,\,\,\,\,\,\,\,\,\,\,\,\,\,\,\,\,\,\,\,\,\,\,\,\,\,\,\,\,\,\,\,\,\,\,{\bf
0}\\\!\!-\,2\,\hat{e}_{0-}\!\cdot\!\bigg[\!{\displaystyle{\frac{\partial\hat{n}}{\partial\vec{r}_{-}}}}\,\hat{n}\,+\,\hat{n}{\displaystyle{\frac{\partial\hat{n}}{\partial\vec{r}_{-}}}}\bigg]\,\,\,\,\,\,\,{\bf
I}\,-\,2\hat{n}\hat{n}\!\end{array}\right]
\label{e24}
\end{eqnarray}
[obtained from Eq. (\ref{e3})]. Starting with the expression of
$\vec{\delta r}_{+}$, we obtain, from Eqs. (\ref{e16}) and
(\ref{e21}-\ref{e24}) that
\begin{eqnarray}
\vec{\delta r}_{+}\,=\,\vec{\delta r}_{-}\,-\,2\,(\vec{\delta
r}_{-}\cdot\hat{n})\,\hat{n}\,,
\label{e25}
\end{eqnarray}
which, together with Eqs. (\ref{e17}-\ref{e19}) yields
\begin{eqnarray}
\hat{e}_{0+}\cdot\vec{\delta
r}_{+}\,=\,0\hspace{0.5cm}{\mbox{and}}\hspace{0.5cm}\delta
r_{i+}\,=\,\delta r_{i-}\,.\hspace{0.5cm}i\,=\,1,\,2
\label{e26}
\end{eqnarray}
In a similar manner, the expression of $\vec{\delta p}_{+}$ can also
be easily obtained from Eq. (\ref{e22}-\ref{e24}), i.e,
\begin{eqnarray}
\vec{\delta p}_{+}\,=\,{\bf A}\cdot\vec{\delta r}^{*}_{-}\,+\,({\bf
I}\,-\,2\hat{n}\hat{n})\,\vec{\delta p}_{-}\,+\,({\bf
I}\,-\,2\hat{n}\hat{n})\,\sum_{i\,=\,1}^{2}(\vec{F}^{*}\cdot\hat{e}_{i-})\,\hat{e}_{i-}\,\delta\tau\,-\,\sum_{i\,=\,1}^{2}(\vec{F}^{*}\cdot\hat{e}_{i+})\,\hat{e}_{i+}\,\delta\tau\,,
\label{e27}
\end{eqnarray}
\noindent where ${\bf
A}\,=\,-\,2\hat{e}_{0-}\cdot\bigg[{\displaystyle{\frac{\partial\hat{n}}{\partial\vec{r}_{-}}}}\,\hat{n}\,+\,\hat{n}{\displaystyle{\frac{\partial\hat{n}}{\partial\vec{r}_{-}}}}\bigg]$
and $\vec{\delta r}^{*}_{-}\,=\,(\vec{\delta
r}_{-}\,+\,\hat{e}_{0-}\,\delta\tau)$ is the infinitesimal vector
$\vec{AC}$ lying on the surface of the scatterer (see Fig. 2).

The simplification of the expression on the r.h.s. of Eq. (\ref{e27})
is carried out in Appendix A. After collecting all the terms together
from Eqs. (\ref{e28}), (\ref{e29}) and (\ref{e33}), we obtain
\begin{eqnarray}
\delta p_{i+}\,=\,\sum_{j\,=\,1}^{2}\bigg[\,\delta_{ij}\,\delta
p_{j-}\,-\,2\,\bigg\{\frac{(\vec{F}^{*}\cdot\hat{n})(\hat{e}_{i-}\cdot\hat{n})(\hat{e}_{j-}\cdot\hat{n})}{(\hat{e}_{0-}\cdot\hat{n})}\,+\,\frac{1}{a}\frac{(\hat{e}_{i-}\cdot\hat{n})(\hat{e}_{j-}\cdot\hat{n})+\,\delta_{ij}\,(\hat{e}_{0-}\cdot\hat{n})^{2}}{(\hat{e}_{0-}\cdot\hat{n})}\bigg\}\,\delta
r_{j-}\bigg]\,.
\label{e34}
\end{eqnarray}

One can now see from Eqs. (\ref{e20}), (\ref{e25}) and (\ref{e34})
that the $4\times4$ matrix ${\bf M}$ has the following structure
\begin{eqnarray}
{\bf M}\,=\,\left[\begin{array}{cc}\,\,{\bf
I}\,\,\,\,\,\,\,\,\,\,\,\,\,\,\,{\bf 0}\\{\bf
R}_{c}\,\,\,\,\,\,\,\,\,\,\,{\bf I}\end{array}\right]\,,
\label{e35}
\end{eqnarray}
where the $2\times2$ matrix ${\bf R}_{c}$ is symmetric. This allows us
to conclude that in the four-dimensional subspace representation, the
transformation of $\delta{\bf\Gamma}$ over a collision is symplectic,
i.e,
\begin{eqnarray}
{\bf M}^{\,\scriptsize{\mbox T}}\,{\bf J}\,{\bf M}\,=\,{\bf J}\,.
\label{e36}
\end{eqnarray}
Finally, using Eqs. (\ref{e4}), (\ref{e15}), (\ref{e36}) and
$\mu(t)=\exp\left[\int_{0}^{t}\alpha(t')dt'\right]$, the
four-dimensional subspace representation, ${\bf L}(t)$ is seen to be
$\mu$-symplectic. The four non-zero Lyapunov exponents of this system
can therefore be arranged in pairs such that the sum of each pair is
exactly equal to $-\langle\alpha\rangle_{t}$, the negative of the long
time average of $\alpha$ along the reference trajectory, i.e., the CPR
is exactly satisfied for this system in the restricted
four-dimensional subspace.

\subsection{Further Generalizations of this Geometric Construction}

We end this section with the following generalizations:\\ {\it (i)}\/
The proof can be extended to any arbitrary dimensions, simply by
including appropriate number of unit vectors $\hat{e}_{i}$'s.\\ {\it
(ii)}\/ Equation (\ref{e34}) demonstrates the role of geometry of the
scatterers and the effect of the external field on the symmetry
property of ${\bf R}_{c}$, as {\it having a symmetric ${\bf R}_{c}$ is
essential for the symplecticity of ${\bf M}$} [and consequently, the
$\mu$-symplecticity of ${\bf L}(t)$]. The effect of the geometry of
the scatterers is coded in the ${\bf A}\cdot\vec{\delta r}^{*}_{-}$
term of Eq. (\ref{e27}). For nonspherical scatterers, this geometric
construction works, so long as the surfaces of the scatterers are
smooth. One has to notice that at the collision point A (see Fig. 1)
on an $(n-1)$-dimensional surface embedded in an $n$-dimensional
Euclidean manifold, $\vec{\delta r}^*_-$ is the infinitesimal vector
$\vec{AC}$ along the surface of the scatterer, and since $\vec{\delta
n}$ appearing in Eq. (\ref{e31}) is the infinitesimal difference
between the two unit normal vectors to the surface of the scatterer at
A and C, for a smooth surface, one can define an $n\times n$ matrix
${\bf B}$, such that $\vec{\delta n}={\bf B}\cdot\vec{\delta r}^*_-$. In
general, the form of the matrix ${\bf B}$ depends on the shape of the
surface of the scatterer at collision point A. As a special case, if
the scatterer is spherical, then ${\bf B}$ can be explicitly constructed
to be the identity matrix times the inverse radius of curvature of the 
sphere. In terms of ${\bf B}$, one then simply needs to obtain the form 
of ${\bf A}\cdot\vec{\delta r}^{*}_{-}$, analogous to Eq. (\ref{e33}), 
given below as
\begin{eqnarray}
{\bf A}\cdot\vec{\delta
r}^{*}_{-}\,=\,2\sum_{i,\,j\,=\,1}^{n}\bigg[(\hat{e}_{i-}\cdot\hat{n})(\hat{e}_{j-}\cdot{\bf
B}\cdot\hat{e}_{0-})\,+\,(\hat{e}_{j-}\cdot\hat{n})(\hat{e}_{i-}\cdot{\bf
B}\cdot\hat{e}_{0-})\nonumber\\&&\hspace{-6cm}-\,(\hat{e}_{0-}\cdot\hat{n})(\hat{e}_{i-}\cdot{\bf
B}\cdot\hat{e}_{j-})\,-\,\frac{\hat{e}_{0-}\cdot{\bf
B}\cdot\hat{e}_{0-}}{\hat{e}_{0-}\cdot\hat{n}}(\hat{e}_{i-}\cdot\hat{n})(\hat{e}_{j-}\cdot\hat{n})\bigg]\,\delta r_{j-}\,\hat{e}_{i+}\,.
\label{eadd1}
\end{eqnarray}
The term in square bracket in Eq. (\ref{eadd1}) is symmetric in $i$
and $j$, which contributes the symmetry of the matrix ${\bf R}_c$ in
Eq. (\ref{e35}). However, the role of external field on the symmetry
of ${\bf R}_{c}$ is a bit more subtle --- the $\vec{F}^*$-dependent
term in Eq. (\ref{e34}) arises due to the fact that {\it the reference
and the adjacent trajectories do not collide the same scatterer at the
same instant, despite the fact that collisions between the particles
and the scatterers are instantaneous}. As Eq. (\ref{e23}) shows, we
have used the fact that both the dynamics of ${\bf\Gamma}_-$ and
${\bf\Gamma}_+$ involve the {\it same}\/ force $\vec{F}^*$, which is
possible {\it only}\/ if the pre- and the post-collisional values of
$\vec{F}^*$ (i.e., appropriate $\vec{F}^*_-$ and $\vec{F}^*_+$) are
the same. One can therefore conclude that this construction can be
generalized to prove CPR for arbitrary (smooth) shapes of the
scatterers in any dimensions, when the forces on the particles due to
the external field depends only the particles' positions.\\ {\it
(iii)}\/ The proof does not depend on the specific locations of the
scatterers, and hence the CPR holds for any arrangement of the
scatterers in space.

\section{Proof of the CPR for hard sphere gas in three dimensions}

In this section, we consider a gas of $N$ identical moving spheres in
three-dimensions. Each of the spheres has a unit mass and is subjected
to an external force that depends {\it only on its position\/}, as
well as a Gaussian thermostat which keeps the total kinetic energy of
the system at a constant value. The spheres interact with each other
by means of binary elastic collisions. During a flight, where no
collision takes place between any two of the spheres, the equations of
motion of the system are
\begin{eqnarray}
\dot{\vec{r}}_{i'}\,=\,\vec{p}_{i'}\,,\hspace{0.5cm}\dot{\vec{p}}_{i'}\,=\,\vec{F}_{i'}\,-\,\alpha{\vec{p}}_{i'}\,,\hspace{0.5cm}i'\,=\,1...N
\label{e38}
\end{eqnarray}
where $\vec{F}_{i'}$ is the external force on the $i'$-th sphere and
$\alpha$ is the coefficient of dynamical friction representing the
isokinetic Gaussian thermostat (hereafter primed indices will always
indicate sphere numbers). The value of $\alpha$ is set in a way that
it keeps the total kinetic energy of the system,
$\displaystyle{\sum_{i'\,=\,1}^{N}}\frac{p^{2}_{i'}}{2}$ constant, i.e,
\begin{eqnarray}
\alpha\,=\,\left[\,\displaystyle{\sum_{i'\,=\,1}^{N}}\vec{F}_{i'}\cdot\vec{p}_{i'}\,\right]\bigg/\left[\,\displaystyle{\sum_{i'\,=\,1}^{N}}p^{2}_{i'}\,\right]\,.
\label{e39}
\end{eqnarray}
As before, we rescale the time such that
$\displaystyle{\sum_{i'\,=\,1}^{N}}p^{2}_{i'}\,=\,1$. At a collision
between the $i'$-th and the $j'$-th spheres, the post-collisional
positions and momenta are related to the pre-collisional values by
\begin{eqnarray}
\vec{r}_{i'+}\,=\,\vec{r}_{i'-}\,,\hspace{1cm}\vec{r}_{j'+}\,=\,\vec{r}_{j'-}\,,\hspace{1cm}\vec{p}_{i'+}\,=\,\vec{p}_{i'-}\,-\,\{(\vec{p}_{i'-}\,-\,\vec{p}_{j'-})\cdot\hat{n}_{i'j'}\}\,\hat{n}_{i'j'}\quad\quad\mbox{and}\nonumber\\&&\hspace{-10cm}\vec{p}_{j'+}\,=\,\vec{p}_{j'-}\,+\,\{(\vec{p}_{i'-}\,-\,\vec{p}_{j'-})\cdot\hat{n}_{i'j'}\}\,\hat{n}_{i'j'}\,,
\label{e40}
\end{eqnarray}
where $\hat{n}_{i'j'}$ is the unit vector along the line joining the
center of the $i'$-th sphere to the $j'$-th sphere at the time of the
collision. Since such a collision is instantaneous, during any such
binary collision process, the positions and momenta of the spheres not
participating in the collision do not change.

We define the $3N$-dimensional vectors $\vec{R}$ and $\vec{P}$,
assembled respectively from the three-dimensional position and
momentum vectors of the $N$ spheres, such that
\begin{eqnarray}
\vec{R}\,=\,\left[\,\vec{r}_{1}\,,\,\,\vec{r}_{2}\,,\ldots,\,\,\vec{r}_{N}\,\right]
\hspace{0.5cm}{\mbox{and}}\hspace{0.5cm}\vec{P}\,=\,\left[\,\vec{p}_{1}\,,\,\,\vec{p}_{2}\,,\ldots,\,\,\vec{p}_{N}\,\right]\,,
\label{e43}
\end{eqnarray}
with $P^{2}\,=\,1$. We also form two more $3N$-dimensional vectors
$\vec{F}$ and $\hat{N}_{i'j'}$: $\vec{F}$ describes the external force
on the spheres, and $\hat{N}_{i'j'}$ is assembled from the unit vector
$\hat{n}_{i'j'}$; i.e,
\begin{eqnarray}
\vec{F}\,=\,\left[\,\vec{F}_{1}\,,\,\,\vec{F}_{2}\,,\ldots,\,\,\vec{F}_{N}\,\right]\hspace{0.5cm}{\mbox{and}}\hspace{0.5cm}\hat{N}_{i'j'}\,=\,\frac{1}{\sqrt{2}}\left[\,\vec{0}\,,\,\,\vec{0}\,,\ldots,\,\,\hat{n}_{i'j'}\,,\ldots,\,\,,\,\,-\,\hat{n}_{i'j'}\,,\ldots,\,\,\vec{0}\,\right]\,,
\label{e44}
\end{eqnarray}
such that the $i'$-th and the $j'$-th entries of $\hat{N}_{i'j'}$
(they are the only non-zero entries) are $\hat{n}_{i'j'}/\sqrt{2}$ and
$-\hat{n}_{i'j'}/\sqrt{2}$ respectively [satisfying the normalization
condition $\hat{N}_{i'j'}\cdot\hat{N}_{i'j'}\,=\,1$]. Equations
(\ref{e38}-\ref{e39}) can now be rewritten [in the same form as
Eqs. (\ref{e1}) and (\ref{e2})] as
\begin{eqnarray}
\dot{\bf
\Gamma}\,=\,\left[\,\dot{\vec{R}}\,,\,\,\,\dot{\vec{P}}\,\right]\,=\,\left[\,\vec{P}\,,\,\,\,\vec{F}\,-\,\alpha\vec{P}\,\right]\,,
\hspace{0.5cm}{\mbox{with}}\hspace{0.5cm}\alpha\,=\,\vec{F}\cdot{\vec{P}}\,,
\label{e45}
\end{eqnarray}
while the collision dynamics can be rewritten as
\begin{eqnarray}
{\bf \Gamma}_{+}\,=\,{\bf Q}\,{\bf
\Gamma}_{-}\,=\,\left[\,\vec{R}_{-}\,,\,\,\,\,\,\vec{P}_{-}\,-\,2\,(\vec{P}_{-}\cdot\hat{N}_{i'j'})\,\hat{N}_{i'j'}\,\right]\,,
\,
\label{e46}
\end{eqnarray}
analogous to Eq. (\ref{e3}). As discussed in the introduction, the
geometric construction associated with Eqs. (\ref{e43}-\ref{e46}) can
be obtained as an extension of the corresponding constructions for the
three-dimensional Lorentz gas. We will see that the construction of
the unit normal vector $\hat{N}_{i'j'}$ is the key component of the
proof of $\mu$-symplecticity and the CPR for hard-sphere gases.

The proof of the CPR proceeds exactly in the same way as it has been
described in section II. The time evolution of the infinitesimal
$6N$-dimensional phase space separation ${\bf\Gamma}(t)$ between the
reference and the adjacent trajectories can be decomposed by means of
$6N$-dimensional ${\bf H}$ and ${\bf M}$ matrices as in
Eq. (\ref{e4}). The $(6N-2)$-dimensional reduction of ${\bf H}$
matrices can be subsequently obtained by constructing $N$
$3N$-dimensional basis vectors
$\hat{e}_{0}(t=0),\,\hat{e}_{1}(t=0),.,.,.,.,\,\hat{e}_{(3N-1)}(t=0)$
at ${\bf\Gamma}_0$, and then parallelly transporting them using
equations analogous to Eqs. (\ref{e8}-\ref{e9}) and choosing to
measure both $\vec{\delta R}$ and $\vec{\delta P}$ in directions
orthogonal to $\hat{e}_{0}$ in the same manner as in
Eqs. (\ref{e7c}-\ref{e7d}). As a special case of what Dettmann and
Morriss considered \cite{DM_pre_96}, in terms of the
$(6N-2)$-dimensional representation, the matrix ${\bf
H}(t_{j}\,-\,t_{j-1})$ is easily seen to be $\mu$-symplectic with
$\mu(t_{j}\,-\,t_{j-1})\,=\,\exp\,\left[\int_{t_{j-1}}^{t_{j}}\alpha(t')dt'\right]$
along the reference trajectory, for a flight between $t_{j\,-\,1}$ and
$t_{j}$. What remains to be shown, therefore, is  that the matrix
${\bf M}$ describing the transformation of $\delta{\bf\Gamma}(t)$ due
to a binary collision is also symplectic. To this end, following
section II of this paper, we use a simple extension, comprising of the
same form of parallel transport (as in Eqs. (\ref{e16}) and
(\ref{e17})) of the $N$ basis vectors
$\hat{e}_{0},\,\hat{e}_{1},\,.\,.\,.\,\hat{e}_{(3N-1)}$ over the
binary collision between the $i'$-th and the $j'$-th sphere, i.e,
\begin{eqnarray}
\hat{e}_{0 +}\,=\,\hat{e}_{0 -}\,-\,2\,(\hat{e}_{0
-}\cdot\hat{N}_{i'j'})\,\hat{N}_{i'j'}\quad\quad\quad\mbox{and}\quad\quad\quad\hat{e}_{i
+}=\,\hat{e}_{i -} - 2\,(\hat{e}_{i
-}\cdot\hat{N}_{i'j'})\,\hat{N}_{i'j'}\,.\hspace{0.3cm}\!1\leq
i\leq\!(3N\!-\!1)
\label{e48}
\end{eqnarray}

Evaluation of the expression of ${\bf M}$ in this $(6N-2)$-dimensional
representation is a bit more involved than that presented in Sec.  II.
Nevertheless, one can still use Fig. 2 to illustrate the collisions at
A and C and the construction of the planes $\Delta_{-}$ and
$\Delta_{+}$ in a schematic way, keeping in mind that this time the
diagram describes quantities in $3N$-dimensions (for another version
of Fig. 2 in the context of hard-sphere gases, the reader may also find
Fig. 6 in \cite{GD_pre_95} helpful). We will follow the logical steps
described in Eqs. (\ref{e21}-\ref{e24}) to obtain the expressions of
$\vec{\delta X}_{+}$ and $\vec{\delta P}_{+}$. Let us first define
below the $6N$ dimensional vector $\delta{\bf \Gamma}_{\pm}$
describing the infinitesimal post(pre)-collisional phase-space
separations of the two trajectories
\begin{eqnarray}
{\bf \delta\Gamma}_{\pm}=\left[\,\vec{\delta
R}_{\pm}\,,\,\,\,\vec{\delta P}_{\pm}\,\right]\,=\,\left[\,\vec{\delta
r}_{1\pm}\,,\ldots,\,\vec{\delta r}_{N\pm}\,,\,\,\,\vec{\delta
p}_{1\pm}\,,\ldots,\,\vec{\delta p}_{N\pm}\,\right]\,=\,
\left[\,\displaystyle{\sum_{i\,=\,1}^{3N-1}}\,\delta
R_{i\pm}\,\hat{e}_{i\pm}\,,\,\,\,\displaystyle{\sum_{i\,=\,1}^{3N-1}}\,\delta
P_{i\pm}\,\hat{e}_{i\pm}\,\right]\,.
\label{e50}
\end{eqnarray}

We observe that the time lag between the binary collisions, involving
the $i'$-th and the $j'$-th sphere on the reference and the adjacent
trajectories (schematically at A and C respectively in Fig. 2) is
\begin{eqnarray}
\delta\tau\,=\,-\,\frac{(\vec{\delta r}_{j'-}\,-\,\vec{\delta
r}_{i'-})\cdot\hat{n}_{i'j'}}{(\vec{p}_{j'-}\,-\,\vec{p}_{i'-})\cdot\hat{n}_{i'j'}}\,=\,-\,\frac{\vec{\delta
R}_{-}\cdot\hat{N}_{i'j'}}{\hat{e}_{0-}\cdot\hat{N}_{i'j'}}\,.
\label{e51}
\end{eqnarray}

Following the procedure outlined in \cite{DPH_pre_96},  we find that
the infinitesimal  phase space separation of the two trajectories at A
and C is, as before,
\begin{eqnarray}
\delta{\bf\Gamma}^{*}\,=\,\,{\bf
\delta\Gamma}_{-}\,+\,\dot{\bf\Gamma}_{-}\,\delta\tau\quad\quad\mbox{and}\quad\quad{\bf
\delta\Gamma}_{+}\,=\,\frac{\partial{\bf Q}}{\partial{\bf
\Gamma}_{-}}\cdot\delta{\bf\Gamma}^{*}\,-\,\dot{\bf\Gamma}_{+}\,\delta\tau\,,
\label{e52}
\end{eqnarray}
where $\dot{\bf\Gamma}_{-}\,(\dot{\bf\Gamma}_{+})$ is obtained from
the equations of motion of the system for the flight immediately
before and after the collision at A (from Eqs. (\ref{e45})), i.e,
\begin{eqnarray}
\dot{\bf
\Gamma}_{\pm}\,=\,\left[\,\hat{e}_{0\pm}\,,\,\,\,\displaystyle{\sum_{i\,=\,1}^{2}}(\vec{F}^{*}\cdot\hat{e}_{i\pm})\,\hat{e}_{i\pm}\,\right]\,.
\label{e54}
\end{eqnarray}
Here $\vec{F}^{*}$ is the force on the reference point, described in
Eq. (\ref{e44}), due to the external field at the point of collision
at A. Using Eq. (\ref{e46}),
\begin{eqnarray}
\frac{\partial{\bf Q}}{\partial{\bf
\Gamma}_{-}}\,=\!\left[\begin{array}{cc}{\,\,\,\,\,\,\,\,\,\,\,\,\,\,\,\,\,\,\,\,\,\,\,\,\,\,\,\,\bf
I}\,\,\,\,\,\,\,\,\,\,\,\,\,\,\,\,\,\,\,\,\,\,\,\,\,\,\,\,\,\,\,\,\,\,\,\,\,\,\,\,\,\,\,\,\,\,\,\,\,\,\,\,\,\,\,\,\,\,\,\,\,\,\,\,\,\,\,\,\,\,\,\,\,\,\,\,\,\,\,{\bf
0}\\\!\!-\,2\,\hat{e}_{0-}\!\cdot\!\bigg[\!{\displaystyle{\frac{\partial\hat{N}_{i'j'}}{\partial\vec{R}_{-}}}}\,\hat{N}_{i'j'}\,+\,\hat{N}_{i'j'}{\displaystyle{\frac{\partial\hat{N}_{i'j'}}{\partial\vec{R}_{-}}}}\bigg]\,\,\,\,\,\,\,\,\,\,\,\,\,{\bf
I}\,-\,2\hat{N}_{i'j'}\hat{N}_{i'j'}\!\end{array}\right]\!,
\label{e55}
\end{eqnarray}
\noindent where each entry of the matrix on the r.h.s. of
Eq. (\ref{e55}) is a $3N\times3N$ matrix.

Starting with the expression of $\vec{\delta R}_{+}$, once again we
obtain, from Eqs. (\ref{e48}) and (\ref{e51}-\ref{e55}) that
\begin{eqnarray}
\vec{\delta R}_{+}\,=\,\vec{\delta R}_{-}\,-\,2\,(\vec{\delta
R}_{-}\cdot\hat{N}_{i'j'})\,\hat{N}_{i'j'}\,,
\label{e56}
\end{eqnarray}
which, together with Eq. (\ref{e48}) yields
\begin{eqnarray}
\hat{e}_{0+}\cdot\vec{\delta R}_{+}\,=\,0\hspace{1cm}
\mbox{and}\hspace{1cm} \delta R_{i+}\!=\delta
R_{i-}.\hspace{0.6cm}1\leq i\leq(3N-1)
\label{e57}
\end{eqnarray}
In a similar manner, the expression of $\vec{\delta P}_{+}$ can also
be easily obtained from Eqs. (\ref{e52}-\ref{e55}), i.e,
\begin{eqnarray}
\vec{\delta P}_{+}\,=\,{\bf A}\cdot\vec{\delta R}^{*}_{-}\,+\,({\bf
I}\,-\,2\hat{N}_{i'j'}\hat{N}_{i'j'})\,\vec{\delta
P}_{-}\,\nonumber\\&&\hspace{-2cm}+\,({\bf
I}\,-\,2\hat{N}_{i'j'}\hat{N}_{i'j'})\,\sum_{i\,=\,1}^{3N-1}(\vec{F}^{*}\cdot\hat{e}_{i-})\,\hat{e}_{i-}\,\delta\tau\,-\,\sum_{i\,=\,1}^{3N-1}(\vec{F}^{*}\cdot\hat{e}_{i+})\,\hat{e}_{i+}\,\delta\tau\,,
\label{e58}
\end{eqnarray}
\noindent where ${\bf
A}\,=\,-\,2\hat{e}_{0-}\cdot\bigg[{\displaystyle{\frac{\partial\hat{N}_{i'j'}}{\partial\vec{R}_{-}}}}\,\hat{N}_{i'j'}\,+\,\hat{N}_{i'j'}{\displaystyle{\frac{\partial\hat{N}_{i'j'}}{\partial\vec{R}_{-}}}}\bigg]$
and $\vec{\delta R}^{*}_{-}\,=\,(\vec{\delta
R}_{-}\,+\,\hat{e}_{0-}\,\delta\tau)$. The simplification of the
expression of $\vec{\delta P}_{+}$ is carried out in Appendix
B. Finally, using Eqs. (\ref{e59}-\ref{e60}) and
(\ref{e61b}-\ref{e61c}), we obtain that in terms of the
$(6N-2)$-dimensional representation, the matrix ${\bf M}$ can be
written in the form
\begin{eqnarray}
\delta P_{i+}\,=\,\sum_{j\,=\,1}^{3N-1}\bigg[\,\delta_{ij}\,\delta
P_{i-}\,+\,\bigg\{W_{ij}\,-\,2\sum_{i,\,j\,=\,1}^{3N-1}\frac{(\vec{F}^{*}\cdot\hat{N}_{i'j'})(\hat{e}_{i-}\cdot\hat{N}_{i'j'})(\hat{e}_{j-}\cdot\hat{N}_{i'j'})}{(\hat{e}_{0-}\cdot\hat{n})}\bigg\}\,\delta
R_{j-}\bigg]
\label{e58a}
\end{eqnarray}
\noindent where $W_{ij}=W_{ji}$ has been evaluated in Eq. (\ref{ae8})
in Appendix B. Equation (\ref{e58a}), together with
Eqs. (\ref{e57}-\ref{e58}) implies that ${\bf M}$ has the same form as
in Eq. (\ref{e35}), where the $(3N-1)\times(3N-1)$ matrix ${\bf
R}_{c}$ is symmetric. This again allows us to conclude that in the
$(6N-2)$-dimensional subspace representation, ${\bf M}$ is symplectic,
and  ${\bf L}(t)$ is $\mu$-symplectic with
$\mu(t)=\exp\left[\int_{0}^{t}\alpha(t')dt'\right]$, i.e., the CPR is
exactly satisfied in the restricted $(6N-2)$-dimensional subspace for
this system.

\subsection{Further Generalizations of this Geometric Construction}

We end this section with the following observations:\\ {\it (i)\/} the
condition that each sphere is of unit mass is not essential --- if the
$i'$-th particle has a mass $m_{i'}$, one can define the new
quantities $\vec{r}\,'_{i'}={m}^{\frac{1}{2}}_{i'}\vec{r}_{i'}$,
$\vec{p}\,'_{i'}={m}^{-\frac{1}{2}}_{i'}\vec{p}_{i'}$ and
$\vec{F}^{\,'}_{i'}={m}^{-\frac{1}{2}}_{i'}\vec{F}_{i'}$ and then
begin with the Eqs. (\ref{e38}-\ref{e40}) replacing $\vec{r}_{i'}$'s,
$\vec{p}_{i'}$'s and $\vec{F}_{i'}$'s by the corresponding primed
variables. One also needs to use
\begin{eqnarray}
\hat{N}^{'}_{i'j'}\,=\,\left[\,\vec{0}\,,\,\,\vec{0}\,,\ldots,\,\,\displaystyle{\sqrt{\frac{m_{j'}}{m_{i'}+m_{j'}}}}\,\hat{n}_{i'j'}\,,\ldots,\,\,,\,\,-\,\displaystyle{\sqrt{\frac{m_{i'}}{m_{i'}+m_{j'}}}}\hat{n}_{i'j'}\,,\ldots,\,\,\vec{0}\,\right]\,,
\label{enote1}
\end{eqnarray}\\ 
such that the $i'$-th and the $j'$-th entries of $\hat{N}^{'}_{i'j'}$
(they are the only non-zero entries) are
$\displaystyle{\sqrt{\frac{m_{j'}}{m_{i'}+m_{j'}}}}\,\hat{n}_{i'j'}$
and
$-\,\displaystyle{\sqrt{\frac{m_{i'}}{m_{i'}+m_{j'}}}}\,\hat{n}_{i'j'}$
respectively (satisfying the normalization condition
$\hat{N}^{'}_{i'j'}\cdot\hat{N}^{'}_{i'j'}\,=\,1$) and define a
corresponding $6N\times6N$ matrix ${\bf U'}$ (see Eqs. (\ref{ae2}) in
Appendix B and its preceding paragraph) such that
\begin{eqnarray}
{\bf U'}_{i'i'}\!=\!\frac{m_{j'}}{m_{i'}}{\bf
U'}_{j'j'}\!=\!-\sqrt{\frac{m_{j'}}{m_{i'}}}\,{\bf I},\,\,\,\,{\bf
U'}_{i'j'}\!={\bf U'}_{j'i'}\!={\bf I};
\label{enote2}
\end{eqnarray}\\     
{\it (ii)\/} the proof can be trivially extended to any arbitrary
dimensions by including appropriate number of unit vectors
$\hat{e}_{i}$'s and \\ {\it (iii)\/} Following point {\it (ii)\/} of
Sec. II.A, it is easy to generalize this construction, and hence the
proof, to arbitrary (smooth) shapes of the particles. One simply needs
to use an analogous matrix ${\bf B}$ such that $\vec{\delta
N}_{i'j'}={\bf B}\cdot\vec{\delta R}^*_-$. Just from Eq. (\ref{e61}),
one then obtains the same equation as Eq. (\ref{eadd1}):
\begin{eqnarray}
{\bf A}\cdot\vec{\delta
R}^{*}_{-}\,=\,2\sum_{i,\,j\,=\,1}^{n}\bigg[(\hat{e}_{i-}\cdot\hat{N}_{i'j'})(\hat{e}_{j-}\cdot{\bf
B}\cdot\hat{e}_{0-})\,+\,(\hat{e}_{j-}\cdot\hat{N}_{i'j'})(\hat{e}_{i-}\cdot{\bf
B}\cdot\hat{e}_{0-})\nonumber\\&&\hspace{-8cm}-\,(\hat{e}_{0-}\cdot\hat{N}_{i'j'})(\hat{e}_{i-}\cdot{\bf
B}\cdot\hat{e}_{j-})\,-\,\frac{\hat{e}_{0-}\cdot{\bf
B}\cdot\hat{e}_{0-}}{\hat{e}_{0-}\cdot\hat{N}_{i'j'}}(\hat{e}_{i-}\cdot\hat{N}_{i'j'})(\hat{e}_{j-}\cdot\hat{N}_{i'j'})\bigg]\,\delta
R_{j-}\,\hat{e}_{i+}\,.
\label{eadd2}
\end{eqnarray}
Notice already the striking similarity of the term inside the square
brackets in Eqs. (\ref{eadd1}) and (\ref{eadd2}) to the form of
$W_{ij}$ in Eq. (\ref{ae8}). When the $nN\times nN$ matrix ${\bf B}$
is decomposed into $N\times N$ blocks of $n\times n$ matrices, the
only non-zero entries are ${\bf B}_{i'i'}$, ${\bf B}_{i'j'}$, ${\bf
B}_{j'i'}$ and ${\bf B}_{j'j'}$, corresponding to the two colliding
particles $i'$ and $j'$. As a special case, if these particles are
spheres, then the matrix ${\bf B}$ can be explicitly constructed to be
${\bf U}/[\sqrt{2}(a_{i'}+a_{j'})]$, where $a_{i'}$ and $a_{j'}$ are
the radii of the colliding particles \cite{PZ_unpub_99} (in Appendix
B, we have used $a_{i'}=a_{j'}=a$). The factor of $\sqrt{2}$ follows
from Eqs. (\ref{e44}) and (\ref{e61a1}). Once again, the form of
Eqs. (\ref{eadd2}) contributes the symmetry of ${\bf R}_c$. In a
similar manner as explained in point {\it (iii)\/} of Sec. II.A, the
symplecticity property of ${\bf M}$ (and hence the $\mu$-symplecticity
of ${\bf L}(t)$ crucially hinges upon the symmetry of ${\bf R}_c$, for
which it is necessary to have the expression of $\vec{F}^*$ to be
invariant under a collision.

\section{Discussion}

While much of the content of this paper is technical, the underlying
principle behind our discussion of the CPR  is similar to those in the
existing literature \cite{WL_cmp_98,Ruelle_jsp_99}. First, the
dimension of the phase space is identified where all the Lyapunov
exponents are non-zero. The dimension of this reduced phase space,
characterized by all the non-zero Lyapunov exponents, depends on the
number of macroscopic conserved quantities in the problem (such as
total momentum, total angular momentum, total energy etc.) that are
consistent with the dynamics. By means of confining the dynamics to a
hypersurface of dimension 1 less than that of the full phase space,
each such conserved quantity reduces the dimension of the phase space
characterized by all the non-zero Lyapunov exponents  by 1. In
addition, the fact that two points in the phase space do not separate
exponentially in time if one point is chosen in the direction of flow
of the other, reduces the dimension of the phase space characterized
by the non-zero Lyapunov exponents also by 1. The dynamics for systems
with hard-core inter-particle interactions are then decomposed into
flights and collisions. The dynamics of a flight segment in the
reduced phase space is shown to be $\mu$-symplectic as a special case
of the explicit geometric construction in Ref. \cite{DM_pre_96}. By
means of another simple explicit geometric construction, it is also
possible to incorporate the collisions in this formalism, which
subsequently leads one to evaluate the matrix ${\bf M}$ and
demonstrate that it is symplectic. Finally, the matrix ${\bf L}(t)$,
assembled from the sequential products of the ${\bf H}$ and ${\bf M}$
matrices, is easily seen to be $\mu$-symplectic using the fact that
the product  of a symplectic and a $\mu$-symplectic transformation is
$\mu$-symplectic.  We must note here that the proof of CPR by this
construction is {\it sufficient and not necessary}; and as a result,
if a system does not satisfy this proof, it is {\it not\/} guaranteed
that CPR would not be satisfied for it. Moreover, this explicit
geometric construction, used in this paper for the parallel transport
of basis vectors, is by no means unique. One could conceivably use a
different set of equations for the parallel transport mechanisms for
both flight and collision parts of the dynamics to prove the CPR for a
dynamical system.

Questions naturally arise about the applicability of this construction
for other hard-particle systems. It turns out that using the same
explicit geometric construction that is presented here, CPR can also
be proved for a gas of hard particles (of finite sizes) in an external
field (such that the forces on the particles due to this field depend
only on their positions of the particles) under a Nos\'e-Hoover
thermostat. One simply has to separate the time evolution of
$\delta{\bf\Gamma}$ in terms of the ${\bf H}$ and the ${\bf M}$
matrices in an appropriately reduced phase space. The
$\mu$-symplecticity of the ${\bf H}$ matrices can be trivially
obtained as a special case of Ref. \cite{DM_pre_97}, where the CPR has
been proved for a system of particles that interact with each other by
means of smooth interparticle potentials. The symplecticity property
of the ${\bf M}$ matrices can also be easily seen to be valid, if one
combines the observation of point {\it (ii)\/} of Sec. III.A with
Eq. (15) of Ref. \cite{DM_pre_97}. We have seen in point {\it (ii)\/}
of Sec. III.A that for  ${\bf M}$ to be symplectic, it is necessary
that $\vec{F}^*$ be invariant under a collision, and it is definitely
the case with Eq. (15) of Ref. \cite{DM_pre_97}.

Clearly, this explicit geometric construction allows one to see the
role of the external field in between collisions (that it is necessary
for the force on the particles due to the external field be dependent
only on their positions), the role of $\vec{F}^*$ and the geometry of
the colliding particles (or scatterers, as the case may be) for it to
prove the CPR. However, these conditions, under which this
construction works to prove the CPR, are rather restrictive in the
context of NEMD studies of the transport quantities of the systems of
physical interest, as they exclude a class of very interesting systems
where the external force may depend on the momenta of the particles
--- for example, when the system is subjected to an additional
external magnetic field. Nevertheless, it has allowed us to prove an
important theoretical result regarding the CPR, for a gas of hard
spheres under shear \cite{PZ_unpub_99}. It is also possible that this
choice of parallel transport can be used for numerical computations to
achieve higher accuracy for the Lyapunov exponents.

We end this paper with the observation that while this construction
uses explicit dependence on the co-ordinate system on a Euclidean
manifold, there exists a (co-ordinate independent) differential
geometric method (for example, in Ref. \cite{WL_cmp_98}) that allows
one to work easily on non-Euclidean manifolds \cite{W_jmpa_00} as
well. How this present construction can be generalized to such
non-Euclidean manifolds is presently unclear, but it remains a
challenging task for the future.

ACKNOWLEDGEMENTS: This paper is dedicated to the 65-th birthday of Bob
Dorfman, of whom the author cherishes many happy memories. The author
wishes to thank Bob Dorfman, Henk van Beijeren and Ramses van Zon for
many useful and motivating discussions regarding this subject. The
work was done and the manuscript was written while the author was a
graduate student under the supervision of Bob Dorfman, and the author
likes to express his gratitude to Bob for his careful and patient
review of this manuscript during that time. Henk van Beijeren provided
the kind hospitality at the University of Utrecht during this and
related work. This work was supported by the research grant of Bob
Dorfman, NSF-PHY-9600428 and that of Henk van Beijeren, FOM, SMC and
the NWO Priority Program of Non-Linear Systems, which are financially
supported by the Nederlandse Organisatie voor Wetenschappelijk
Onderzoek (NWO).

\appendix

\section{}

To simplify the expression of $\vec{\delta p}_{+}$, we break the
r.h.s. of Eq. (\ref{e27}) into several parts. First we notice that
\begin{eqnarray}
({\bf I}\,-\,2\hat{n}\hat{n})\,\vec{\delta
p}_{-}\,=\,\sum_{i\,=\,1}^{2}\delta
p_{i-}\,\hat{e}_{i+}\quad\quad\quad\mbox{and}
\label{e28}
\end{eqnarray}
\begin{eqnarray}
({\bf
I}\,-\,2\hat{n}\hat{n})\,\sum_{i\,=\,1}^{2}(\vec{F}^{*}\cdot\hat{e}_{i-})\,\hat{e}_{i-}\,\delta\tau\,-\,\sum_{i\,=\,1}^{2}(\vec{F}^{*}\cdot\hat{e}_{i+})\,\hat{e}_{i+}\,\delta\tau\,=\,-\,2\sum_{i,\,j\,=\,1}^{2}\frac{(\vec{F}^{*}\cdot\hat{n})(\hat{e}_{i-}\cdot\hat{n})(\hat{e}_{j-}\cdot\hat{n})}{(\hat{e}_{0-}\cdot\hat{n})}\,\delta
r_{j-}\,\hat{e}_{i+}\,.
\label{e29}
\end{eqnarray}
Next, we observe that the rest of the r.h.s. of Eq. (\ref{e27}), i.e,
${\bf A}\cdot\,\vec{\delta r}^{*}_{-}$ describes the effect of the
orientation of $\hat{n}$ on $\vec{\delta p}_{+}$. Having  denoted the
unit vector normal to the surface of the scatterer at C by
$\hat{n}^{'}$, which can be related to $\hat{n}$ by
\begin{eqnarray}
\hat{n}'\,=\,\hat{n}\,+\,\vec{\delta n}\,,
\label{e30}
\end{eqnarray}
($\hat{n}\cdot\vec{\delta n}\,=\,0$), it is easily seen from
Eqs. (\ref{e21a}-\ref{e22}) and (\ref{e24}) that
\begin{eqnarray}
{\bf A}\cdot\vec{\delta
r}^{*}_{-}\,=\,-\,2\,[\,(\hat{e}_{0-}\cdot\vec{\delta
n})\,\hat{n}\,+\,(\hat{e}_{0-}\cdot\hat{n})\,\vec{\delta n}]\,.
\label{e31}
\end{eqnarray}
Furthermore, using $\vec{\delta n}=\vec{\delta r}^{*}_{-}/a$, where
$a$ is the radius of the scatterer, The expression of ${\bf
A}\cdot\vec{\delta r}^{*}_{-}$ can be readily simplified as

\begin{eqnarray}
{\bf A}\cdot\vec{\delta
r}^{*}_{-}\,=\,-\,\frac{2}{a}\sum_{i,\,j\,=\,1}^{2}\frac{(\hat{e}_{i-}\cdot\hat{n})(\hat{e}_{j-}\cdot\hat{n})+\,\delta_{ij}\,(\hat{e}_{0-}\cdot\hat{n})^{2}}{(\hat{e}_{0-}\cdot\hat{n})}\,\delta
r_{j-}\,\hat{e}_{i+}\,.
\label{e33}
\end{eqnarray}

\section{}

Two terms on the r.h.s. of Eq. (\ref{e58}) simplify as before, i.e,
\begin{eqnarray}
({\bf I}\,-\,2\hat{N}_{i'j'}\hat{N}_{i'j'})\,\vec{\delta
P}_{-}\,=\,\sum_{i\,=\,1}^{3N-1}\delta
P_{i-}\,\hat{e}_{i+}\quad\quad\quad\mbox{and}
\label{e59}
\end{eqnarray}
\begin{eqnarray}
({\bf
I}\,-\,2\hat{N}_{i'j'}\hat{N}_{i'j'})\,\sum_{i\,=\,1}^{3N-1}(\vec{F}^{*}\cdot\hat{e}_{i-})\,\hat{e}_{i-}\,\delta\tau\,-\,\sum_{i\,=\,1}^{3N-1}(\vec{F}^{*}\cdot\hat{e}_{i+})\,\hat{e}_{i+}\,\delta\tau\nonumber\\&&{\hspace{-3.5cm}}=\,-\,2\sum_{i,\,j\,=\,1}^{3N-1}\frac{(\vec{F}^{*}\cdot\hat{N}_{i'j'})(\hat{e}_{i-}\cdot\hat{N}_{i'j'})(\hat{e}_{j-}\cdot\hat{N}_{i'j'})}{(\hat{e}_{0-}\cdot\hat{n})}\,\delta
R_{j-}\,\hat{e}_{i+}\,.
\label{e60}
\end{eqnarray}
\noindent Following section II, the term ${\bf A}\cdot\vec{\delta
R}^{*}_{-}$ can be expressed as
\begin{eqnarray}
{\bf A}\cdot\vec{\delta
R}^{*}_{-}=-\,2\,[\,(\hat{e}_{0-}\!\cdot\vec{\delta
N}_{i'j'})\,\hat{N}_{i'j'}+(\hat{e}_{0-}\cdot\hat{N}_{i'j'})\,\vec{\delta
N}_{i'j'}]\,,
\label{e61}
\end{eqnarray}
where
\begin{eqnarray}
\vec{\delta{N}}_{i'j'}\,=\,\frac{1}{\sqrt{2}}\left[\,\vec{0}\,,\,\,\vec{0}\,,\ldots,\,\,\vec{\delta{n}}_{i'j'}\,,\ldots,\,\,,\,\,-\,\vec{\delta{n}}_{i'j'}\,,\ldots,\,\,\vec{0}\,\right]\,,
\label{e61a1}
\end{eqnarray}
satisfying $\hat{n}_{i'j'}\cdot\vec{\delta n}_{i'j'}\,=\,0$.  This
orthogonality condition between $\hat{n}_{i'j'}$ and $\vec{\delta
n}_{i'j'}$ also implies that $\hat{N}_{i'j'}\cdot\vec{\delta
N}_{i'j'}\,=\,0$; from which it can be readily shown that
\begin{eqnarray}
\hat{e}_{0+}\cdot[\,{\bf A}\cdot\vec{\delta R}^{*}_{-}]\,=\,0\,,
\label{e61a}
\end{eqnarray}
In other words, we have
\begin{eqnarray}
{\bf A}\cdot\vec{\delta R}^{*}_{-}\,=\,\sum_{i\,=\,1}^{3N-1}\{[\,{\bf
A}\cdot\vec{\delta R}^{*}_{-}]\cdot\hat{e}_{i+}\}\,\hat{e}_{i+}\,.
\label{e61b}
\end{eqnarray}
The full simplification of the expression $[\,{\bf A}\cdot\vec{\delta
R}^{*}_{-}]\cdot\hat{e}_{i+}$ is carried out below, where it is shown
that
\begin{eqnarray}
[\,{\bf A}\cdot\vec{\delta
R}^{*}_{-}]\cdot\hat{e}_{i+}\,=\,\sum_{j\,=\,1}^{3N-1}W_{ij}\,\vec{\delta
R}_{j}\,,
\label{e61c}
\end{eqnarray}
with the property that $W_{ij}\,=\,W_{ji}$ (see Eq. (\ref{ae8})).  The
expression ${\bf A}\cdot\vec{\delta R}^{*}_{-}$, using Eq.
(\ref{e61}), can be written in the following form
\begin{eqnarray}
{\bf A}\cdot\vec{\delta
R}^{*}_{-}\,=\,\left[\,\vec{0}\,,\,\,\vec{0}\,,\ldots,\,\,\vec{A}_{i'}\,,\ldots,\,\,,\,\,-\,\vec{A}_{j'}\,,\ldots,\,\,\vec{0}\,\right]\,;
\label{e62}
\end{eqnarray}
such that all but the $i'$-th and the $j'$-th entries on the r.h.s. of
Eq.(\ref{e62}) are zero, while the $j'$-th entry $\vec{A}_{j'}$ is
related to the $i'$-th entry $\vec{A}_{i'}$ by [using
Eq. (\ref{e61a1})]
\begin{eqnarray}
\vec{A}_{j'}\,=\,-\,\vec{A}_{i'}\,=\,\,[\{(\vec{p}_{i'-}\,-\,\vec{p}_{j'-})\cdot\vec{\delta
n}_{i'j'}\}\,\hat{n}_{i'j'}\,+\,\{(\vec{p}_{i'-}\,-\,\vec{p}_{j'-})\cdot\hat{n}_{i'j'}\}\,\vec{\delta
n}_{i'j'}]\,.
\label{e63}
\end{eqnarray}
To obtain an expression for $\vec{\delta n}_{i'j'}$ analogous to
$\vec{\delta n}=\vec{\delta r}^{*}_{-}/a$ as used in Appendix A, we
need to take a look at Figs. 3 and 4. Figure 3 describes, in the
laboratory frame, the binary collision  process between the $i'$-th
and the $j'$-th sphere on the reference  and adjacent trajectories;
the thick-lined spheres are on the reference trajectory whereas the
thin-lined spheres are on the adjacent trajectory. Figure 4 describes
the same binary collision process in the reference frame of the
$i'$-th sphere (with center C). In Fig. 3,  the thick-lined $j'$-th
sphere (with center D) on the left depicts the collision situation on
the reference trajectory and the thin-lined $j'$-th sphere (with
center E) on the left depicts the collision situation on the adjacent
trajectory. Clearly, in Fig. 4, the  infinitesimal vector $\vec{DE}$
is given by
\begin{eqnarray}
\vec{\delta r}^{*}_{i'j'}\,=\,\vec{\delta r}_{j'-}\,-\,\vec{\delta
r}_{i'-}\,+\,(\vec{p}_{j'-}\,-\,\vec{p}_{i'-})\,\delta\tau
\label{e64}
\end{eqnarray}
and since the lengths of both the lines CD and CE are $2a$ ($a$ is the
radius of each sphere), we have
\begin{eqnarray}
\vec{\delta n}_{i'j'}\,=\,\frac{1}{2a}\,\vec{\delta r}^{*}_{i'j'}\,.
\label{e65}
\end{eqnarray}
Starting with the expression in Eq. (\ref{e61b}), we proceed to
calculate the quantity $[\,{\bf A}\cdot\vec{\delta
R}^{*}_{-}]\cdot\hat{e}_{i+}$. Let us first define a $3N\times3N$
matrix ${\bf U}$ composed of $N\times N$ blocks of $3\times3$
matrices, such that, in terms of the block indices the only non-zero
entries of ${\bf U}$ are
\begin{eqnarray}
{\bf U}_{i'i'}\,=\,-\,{\bf U}_{i'j'}\,=\,-\,{\bf U}_{j'i'}\,=\,{\bf
U}_{j'j'}\,=\,-\,{\bf I}\,,
\label{ae2}
\end{eqnarray}
where ${\bf I}$ is the $3\times3$ unit matrix. The matrix ${\bf U}$
has  the property that
\begin{eqnarray}
{\bf U}^{\scriptsize{\mbox{T}}}\,=\,{\bf
U}\hspace{0.5cm}{\mbox{and}}\hspace{0.5cm}{\bf
U}\cdot\hat{N}_{i'j'}\,=\,-\,2\,\hat{N}_{i'j'}\,.
\label{ae3}
\end{eqnarray}
Using Eqs. (\ref{e62}-\ref{e65}) and (\ref{ae3}), one can then
simplify $[\,{\bf A}\cdot\vec{\delta R}^{*}_{-}]\cdot\hat{e}_{i+}$ as
\begin{eqnarray}
[\,{\bf A}\cdot\vec{\delta
R}^{*}_{-}]\cdot\hat{e}_{i+}\,=\,\frac{\sqrt{2}}{2a}\,(\hat{e}_{i-}\cdot\hat{N}_{i'j'})\sum_{j\,=\,1}^{3N-1}\bigg[\,(\hat{e}_{j-}\cdot{\bf
U}\cdot\hat{e}_{0-})\,-\,(\hat{e}_{0-}\cdot{\bf
U}\cdot\hat{e}_{0-})\,\frac{(\hat{e}_{j-}\cdot\hat{N}_{i'j'})}{(\hat{e}_{0-}\cdot\hat{N}_{i'j'})}\bigg]\,\delta
R_{j-}\,\nonumber\\&&\hspace{-10.8cm}-\,\frac{\sqrt{2}}{2a}\,(\hat{e}_{0-}\cdot\hat{N}_{i'j'})\sum_{j\,=\,1}^{3N-1}\bigg[\,(\hat{e}_{i+}\cdot{\bf
U}\cdot\hat{e}_{j-})\,-\,(\hat{e}_{i+}\cdot{\bf
U}\cdot\hat{e}_{0-})\,\frac{(\hat{e}_{j-}\cdot\hat{N}_{i'j'})}{(\hat{e}_{0-}\cdot\hat{N}_{i'j'})}\bigg]\,\delta
R_{j-}\,.
\label{ae4}
\end{eqnarray}
Equation (\ref{ae4}) can be further simplified by using
\begin{eqnarray}
(\hat{e}_{i+}\cdot{\bf
U}\cdot\hat{e}_{0-})\,(\hat{e}_{j-}\cdot\hat{N}_{i'j'})\,=\,\{(\hat{e}_{i-}\cdot{\bf
U}\cdot\hat{e}_{0-})\,+\,4\,(\hat{e}_{i-}\cdot\hat{N}_{i'j'})\,(\hat{e}_{0-}\cdot\hat{N}_{i'j'})\}\,(\hat{e}_{j-}\cdot\hat{N}_{i'j'})\,,
\label{ae5}
\end{eqnarray}
\begin{eqnarray}
(\hat{e}_{i+}\cdot{\bf
U}\cdot\hat{e}_{j-})\,(\hat{e}_{0-}\cdot\hat{N}_{i'j'})\,=\,\{(\hat{e}_{i-}\cdot{\bf
U}\cdot\hat{e}_{j-})\,+\,4\,(\hat{e}_{i-}\cdot\hat{N}_{i'j'})\,(\hat{e}_{j-}\cdot\hat{N}_{i'j'})\}\,(\hat{e}_{0-}\cdot\hat{N}_{i'j'})
\label{ae6}
\end{eqnarray}
\noindent and Eq. (\ref{ae3}), to obtain
\begin{eqnarray}
[\,{\bf A}\cdot\vec{\delta
R}^{*}_{-}]\cdot\hat{e}_{i+}\,=\,\sum_{j\,=\,1}^{3N-1}W_{ij}\,\delta
R_{j-}\,,
\label{ae7}
\end{eqnarray}
where
\begin{eqnarray}
W_{ij}\,=\,W_{ji}\,=\,\frac{1}{\sqrt{2}a}\,\bigg[(\hat{e}_{i-}\cdot\hat{N}_{i'j'})\,(\hat{e}_{j-}\cdot{\bf
U}\cdot\hat{e}_{0-})\,+\,(\hat{e}_{j-}\cdot\hat{N}_{i'j'})\,(\hat{e}_{i-}\cdot{\bf
U}\cdot\hat{e}_{0-})\,-\,(\hat{e}_{i-}\cdot{\bf
U}\cdot\hat{e}_{j-})\,(\hat{e}_{0-}\cdot\hat{N}_{i'j'})\,\nonumber\\&&\hspace{-8cm}-\,(\hat{e}_{j-}\cdot\hat{N}_{i'j'})\,(\hat{e}_{i-}\cdot\hat{N}_{i'j'})\,\frac{(\hat{e}_{0-}\cdot{\bf
U}\cdot\hat{e}_{0-})}{(\hat{e}_{0-}\cdot\hat{N}_{i'j'})}\bigg]\,.
\label{ae8}
\end{eqnarray}
\noindent Equation (\ref{ae8}) is then used in Eq. (\ref{e61c}).

\eject
\begin{figure}
\epsfxsize=3in
\centerline{\epsfbox{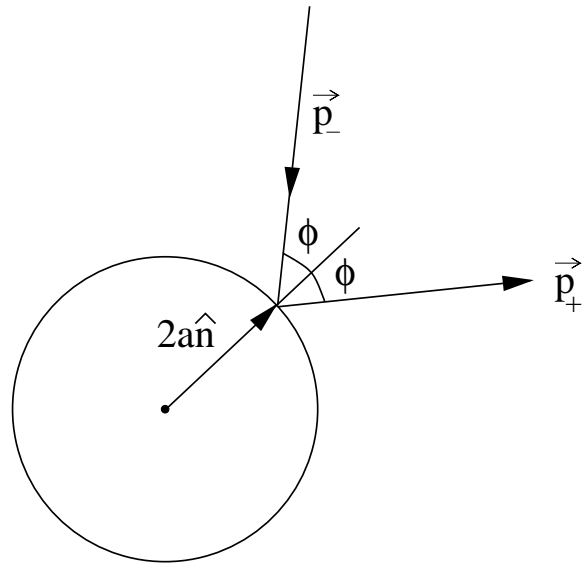}}
\caption{Dynamics of the point particle at a collision in 
two-dimensional projection.}
\end{figure}
 
\begin{figure}
\epsfxsize=5in
\centerline{\epsfbox{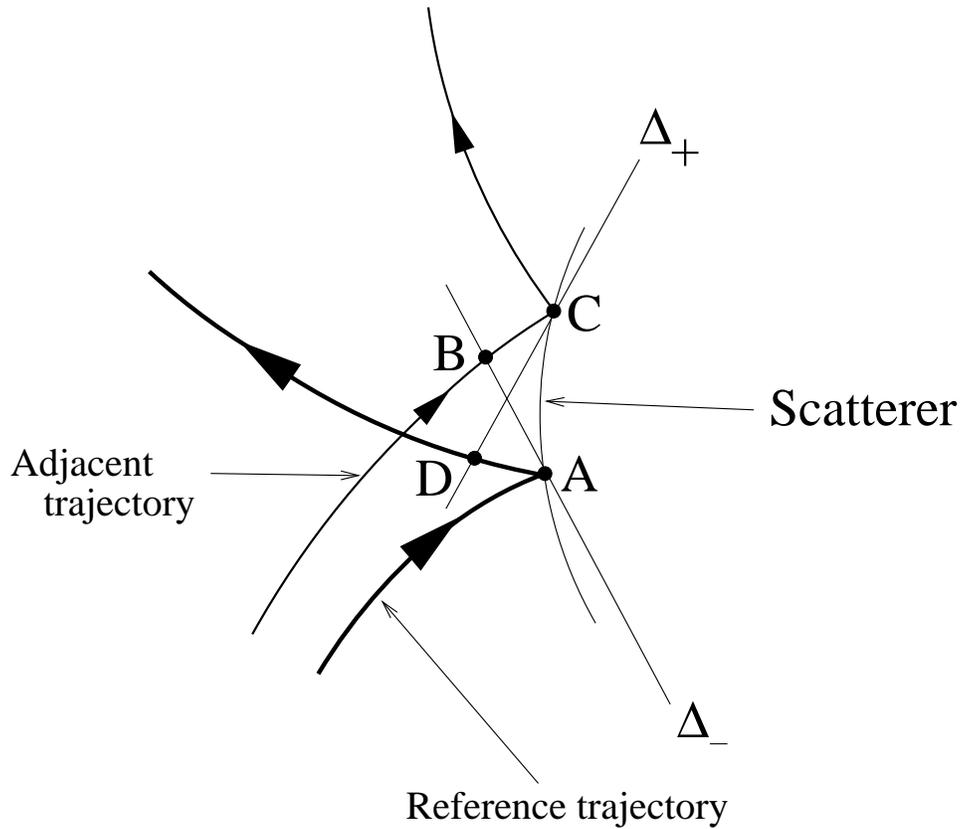}}
\caption{A two-dimensional projection of the three-dimensional
collision dynamics of the reference and adjacent trajectories.}
\end{figure}

\begin{figure}
\epsfxsize=5in
\centerline{\epsfbox{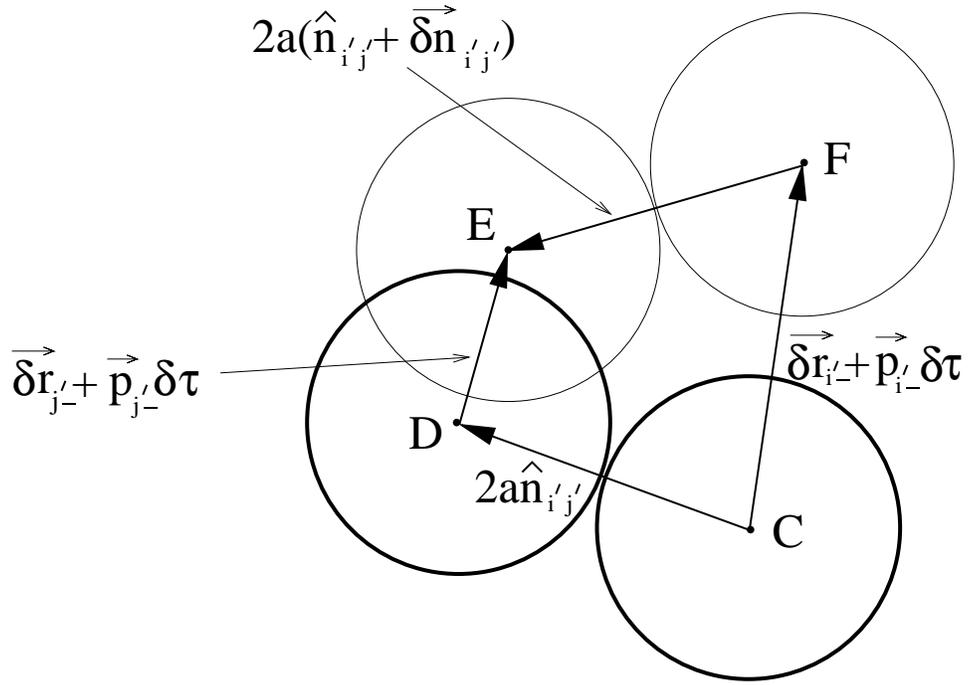}}
\caption{Collision between the $i'$-th and the $j'$-th sphere on the
reference and adjacent trajectories in the laboratory frame. Thick-lined
spheres are on the reference trajectory whereas the thin-lined spheres are
on the adjacent trajectory.}
\end{figure}
 
\begin{figure}
\epsfxsize=5in
\centerline{\epsfbox{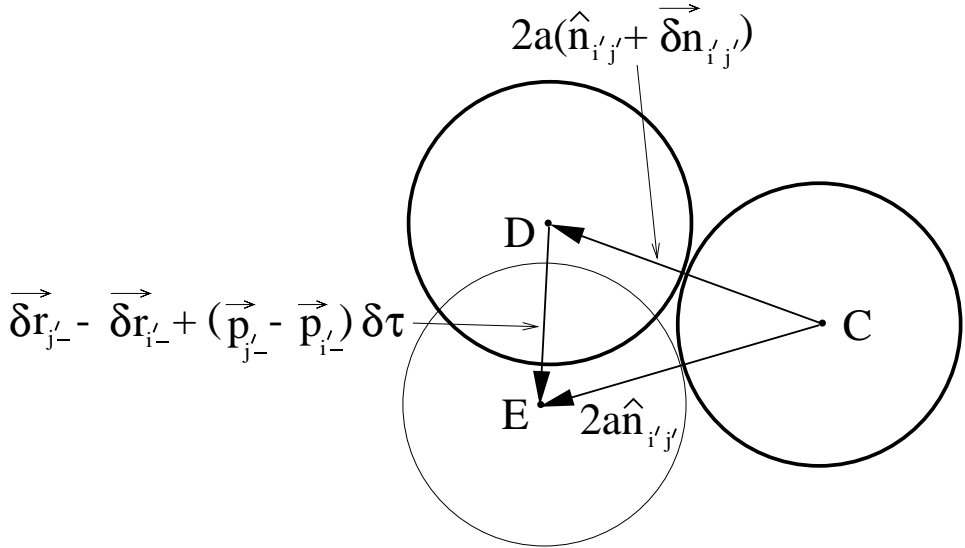}}
\caption{Same collisions as in Fig. 3, in the reference frame of the
$i'$-th sphere.} 
\end{figure}

\end{document}